# High spin states of cation vacancies in GaP, GaN, AlN, BN, ZnO and BeO: A first principles study

O. Volnianska[1] and P. Boguslawski[1,2,*]
1. *Institute of Physics PAS, al. Lotnikow 32/46, 02-668 Warsaw, Poland*
2. *Institute of Physics, University of Bydgoszcz, Weyssenhoffa 11, 85-072 Bydgoszcz, Poland*

High spin states of cation vacancies in GaP, GaN, AlN, BN, ZnO and BeO were analyzed by first principles calculations. The spin-polarized vacancy-induced level is located in the band gap in GaP, ZnO and BeO. In the nitrides, the stronger exchange coupling forces the vacancy states to be resonant with valence bands, forbids formation of positively charged vacancies in GaN and BN, and allows Al vacancy in *p*-AlN to assume the highest possible *S*=2 spin state. The shape of the spin density, isotropic in the zinc blende structure, has a pronounced directional character in the wurtzite structure. Stability of spin polarization of the vacancy states is determined by spin polarization energies of anions, as well as by interatomic distances between the vacancy neighbors, and thus is given by both the lattice constant of the host and the atomic relaxations around the vacancy. Implications for experiment are discussed.

## I. INTRODUCTION

In recent years, magnetism without transition metal ions, thus based on *p* rather than *d* or *f* orbitals, was discovered in a wide variety of systems [1]. The effect originates in the high spin polarization energy of *p* orbitals of light atoms from the second row of the periodic table. The studied systems include ideal crystals that contain $O_2$ or $N_2$ molecules as structural units, *e.g.*, low-temperature phases of solid oxygen, $Rb_4O_6$, and SrN. Room temperature ferromagnetism induced by non-magnetic dopants was first observed in ZnO with a few per cent of C [2]. In the case of both pristine [3] and proton-irradiated [4] graphite, graphene [5] and nominally undoped ZnO [6], the effect was associated with the presence of partially occupied dangling bonds generated by structural defects, such as grain boundaries, nanoribbon edges or vacancies (*V*s). Calculations performed for CaO suggested that magnetic coupling between spins of Ca vacancies can lead to a collective spin polarization [7]. This idea was employed to predict or explain ferromagnetism in other materials [8], but it also was questioned based on the fact that equilibrium concentrations of vacancies are much too low to result in magnetism of as-grown crystals [9]. On the other hand, the local defect concentration in irradiated samples can be sufficient to exceed the percolation limit and achieve ferromagnetism.

An important aspect of this 'physics of dangling bonds' are high spin states of vacancies in semiconductors. Such configurations take place when the vacancy-induced multiplet level is partially occupied by electrons with parallel spins and, depending on the actual occupation, the total spin is 1, 3/2 or 2 rather than 0 or 1/2. Thus, high spin configurations correspond to the local spin polarization of electrons residing at the dangling bonds of the vacancy neighbors. Since various mechanisms of magnetic interactions scale as the spin squared, high spin states guarantee an enhanced *V-V* coupling. The first experimental discovery of a vacancy in a high spin *S*=1 state was achieved for a neutral zinc vacancy $V_{Zn}$ in electron irradiated ZnO [10]. High spin states of cation vacancies $V_{cation}$ were observed also in MgO [11], GaP [12], and both zinc blende [13] and 4H [14] polytypes of SiC. Theoretically, stability of high spin configurations was first analyzed for $V_{Si}$ in SiC [15-18]. Subsequent first principles calculations predicted stability of high spin state of neutral $V_{cation}$ in CaO [7,9], $HfO_2$ [9,19,20], III-V nitrides [21-25], MgO [24,26], ZnO [22,25,27,28] and in II-VI compounds [29]. In this paper we study high spin states of $V_{cation}$ by first-principles calculations [30,31]. We begin with the series GaAs-GaP-GaN of gallium compounds, and show that high spin states originate in the strong spin polarization of the 2*p* shell of light atoms, similarly to the case of bulk compounds [1]. The analysis for III-V nitrides and II-VI oxides reveals previously unrecognized features, such as the resonant character of the vacancy levels in the nitrides, their dependence on the charge state, correlation between the crystal structure and the topology of the spin density, and the impact of the atomic configuration of the vacancy on its electronic structure.

## II. METHOD OF CALCULATIONS

Calculations were done within the density functional theory using the generalized gradient approximations [30], and Vanderbilt pseudopotentials [31] as implemented in the Quantum Espresso code [32]. The orbitals that were chosen as valence orbitals for pseudopotentials are (2*s*, 2*p*) for Be, B, N, and O, (3*s*, 3*p*) for Al and P, (3*d*, 4*s*, 4*p*) for Ga, (4*s*, 4*p*) for As, and (3*d*, 4*s*) for Zn. The cutoff energies for the plane wave basis set were 15 Ry for GaP and GaAs, and 30 Ry for the remaining systems. The used ideal supercells contained 512 atoms for the zinc blende (*zb*) and 128 atoms for the wurtzite (*w*) structure, and the corresponding Brillouin-zone summations were performed using the Γ point and the 2×2×2 k-point mesh [33]. The convergence with respect to the size of the unit cell is briefly addressed in Sec. III. Optimization of ionic positions was stopped when forces acting on ions were smaller than 0.026 eV/Å. The calculated lattice constants in Å for the wurtzite BN, AlN, GaN, BeO and ZnO are ($a$=2.55, $c$=4.2), ($a$=3.12, $c$=5.0), ($a$=3.22, $c$=5.25), ($a$=2.70, $c$=4.38); ($a$=3.25, $c$=5.20), respectively.

## III. NEUTRAL CATION VACANCIES

In both the zinc blende and the wurtzite structure with the tetrahedral coordination of atoms, four $sp^3$ orbitals of four vacancy neighbors combine into a singlet $a_1$ and a triplet $t_2$ that is higher in energy. Energies of $a_1$ and $t_2$ relative to the valence band top (VBT) depend on the crystal, but typically the $a_1$ level of $V_{cation}$ is a resonance degenerate with valence bands, and the triplet $t_2$ is in the band gap. In the $w$ structure, the hexagonal crystal field splits $t_2$ into a doublet $e_2$ and a singlet $a_1$ with a small splitting energy of about 0.2 eV. $t_2$ of a neutral $V_{cation}$ is occupied with 3 and 4 electrons in III-V and II-VI crystals, respectively. In the case of non-vanishing spin polarization, the exchange coupling splits $t_2$ into spin-up $t_{2\uparrow}$ and spin-down $t_{2\downarrow}$ states by $\Delta\varepsilon_{ex} = \varepsilon(t_{2\downarrow}) - \varepsilon(t_{2\uparrow})$, where $\varepsilon$ is the energy of the defect level, and $a_1$ into $a_{1\uparrow}$ and $a_{1\downarrow}$. Stability of various spin configurations is given by spin polarization energy $\Delta E^{pol}$ defined as the difference in total energy of the spin-nonpolarized and high spin states. High spin state is stable if $\Delta E^{pol} > 0$.

### A. III-V compounds

The calculated energy levels of neutral cation vacancies are shown in Fig. 1. With the decreasing anion's atomic number in the series GaAs-GaP-GaN, the spin splitting $\Delta\varepsilon_{ex}$ of the $t_2$ triplet increases due to the increasing strength of exchange coupling of the $p$(anion) shell [1]. More specifically, in GaAs both spin polarization and $\Delta\varepsilon_{ex}$ vanish. Next, in agreement with experiment [12] the $S$=3/2 high-spin state of $V_{Ga}$ in GaP is stable, with $\Delta E^{pol}$= 0.06 eV and $\Delta\varepsilon_{ex}$ =0.19 eV. Stability of the $S$=3/2 state is much more pronounced in III-V nitrides. In the series $zb$-GaN, $zb$-AlN and $zb$-BN, $\Delta E^{pol}$ is 0.69, 0.93 and 0.82 eV, respectively, $i.e.$, it is an order of magnitude higher, and $\Delta\varepsilon_{ex}$ is about 4 times larger than in GaP, see Fig. 1 and Table I. Importantly, the high $\Delta\varepsilon_{ex}$ qualitatively changes electronic structure of vacancies: the $t_{2\uparrow}$ triplet is not a gap state (like in GaP) but a resonance degenerate with valence bands. This previously unrecognized feature changes the vacancy properties, as discussed below.

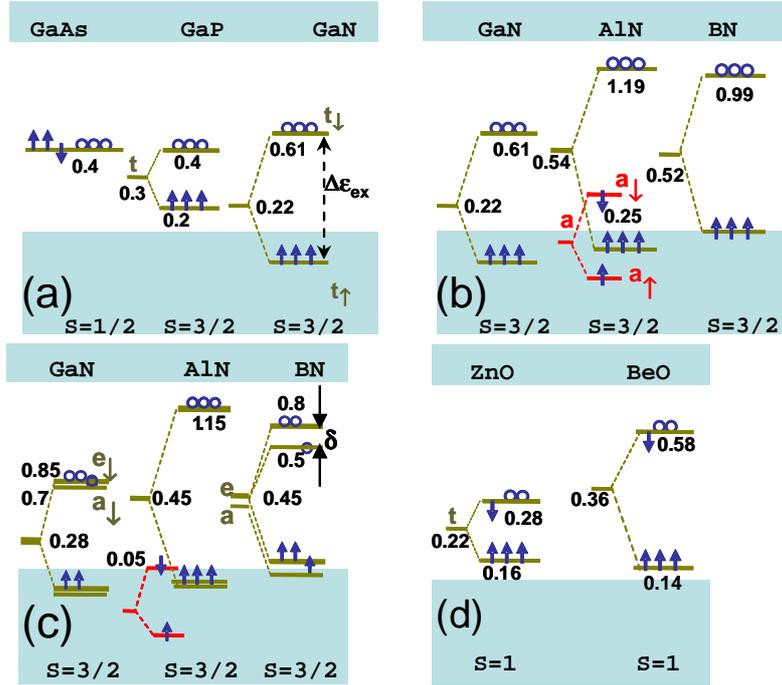

FIG. 1 (color online). Energy levels of a neutral $V_{cation}$ (a) in the series of Ga pnictides GaAs, GaP and GaN, and in the nitrides GaN, AlN and BN with (b) zinc-blende and (c) wurtzite structure, and (d) in $w$-ZnO and $w$-BeO. $t$, $e$ and $a$ denote the $t_2$, $e_2$ and $a_1$ states, respectively, and the hexagonal splitting $\delta$ of $t_2$ is shown for BN. The numbers give the calculated energies relative to VBT in eV. Energies of the resonance states are shown only schematically. Spins of electrons are indicated by arrows, empty spheres indicate unoccupied states.

TABLE I. Polarization energies $\Delta E^{pol}$, exchange splitting $\Delta\varepsilon_{ex}$, and energies of the $t_2$ state for both $zb$ and $w$ structures. All values are in eV. In the cases when $t_{2\uparrow}$ is degenerate with the valence band and its energy cannot be assessed $\Delta\varepsilon_{ex}$ is defined as $2[\varepsilon(t_{2\downarrow})-\varepsilon(t_2)]$.

|  | $\Delta E^{pol}$ | $\Delta\varepsilon_{ex}$ | $t_2$ |
|---|---|---|---|
| $V_{Ga}$ ($zb/w$) | 0.69/0.64 | 0.78/1.1 | 0.22/0.25 |
| $V_{Al}$ ($zb/w$) | 0.93/0.85 | 1.30/1.4 | 0.54/0.45 |
| $V_B$ ($zb/w$) | 0.82/0.55 | 0.94/0.95 | 0.52/0.50 |
| $V_{Zn}$ ($w$) | 0.04 | 0.11 | 0.22 |
| $V_{Be}$ ($w$) | 0.20 | 0.44 | 0.36 |

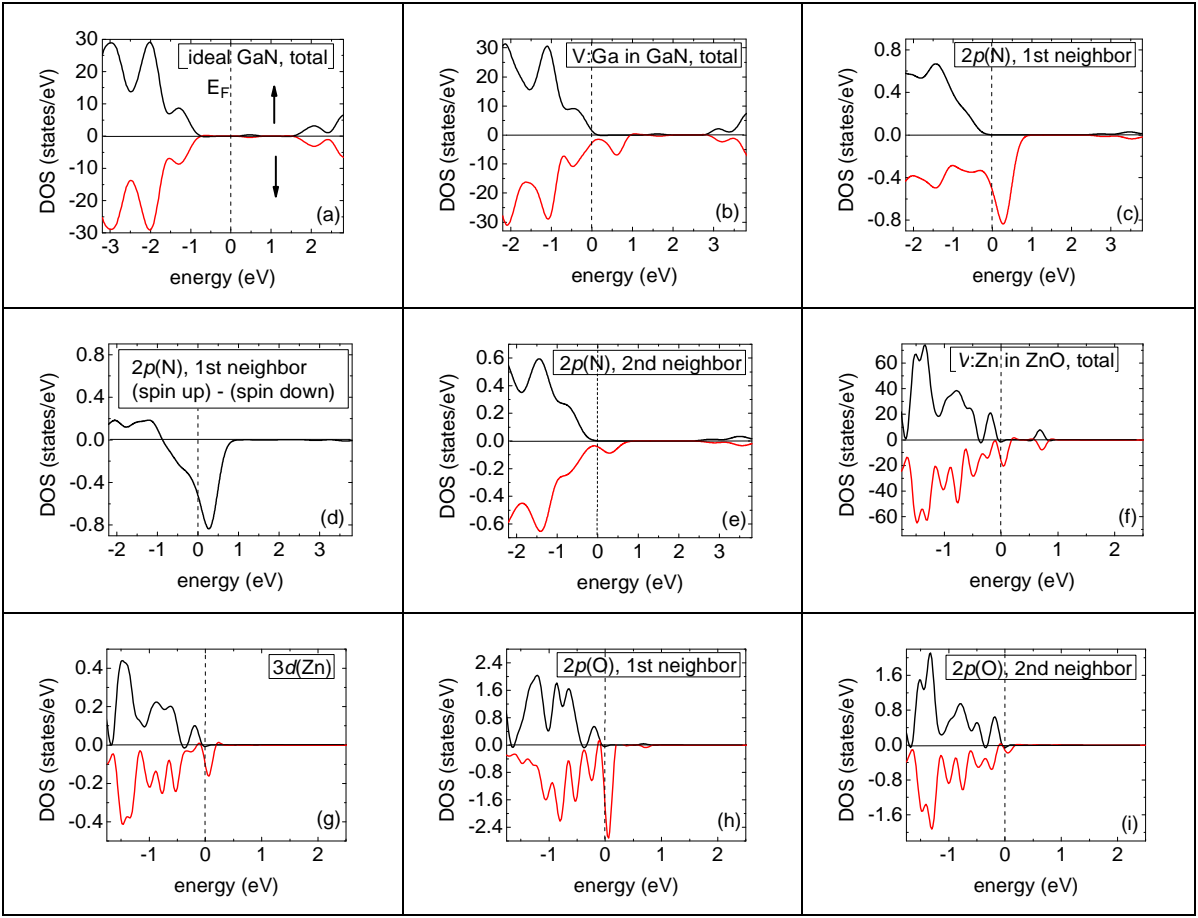

FIG. 2 (color online). Spin-resolved density of states of (a) perfect GaN, (b) GaN with a vacancy, (c) the contribution of $p$(N) orbitals of the first vacancy N neighbor, (d) is the difference of spin-up and spin-down contributions of the first vacancy neighbor, (e) the contribution of $p$(N) orbitals of the second vacancy N neighbor. The results are for 64-atom cells. The zero of energy is at the Fermi energy, denoted by vertical dotted lines. Figure (f) shows the spin-resolved total DOS of ZnO:$V$, and the contribution of $d$(Zn) (g), and $p$(O) of the first (h) and second (i) oxygen neighbors. Negative values denote the spin down channel.

To verify the convergence of the results with respect to the size of the supercell the calculations were performed for cells with 64, 216, and 512 atoms for the $zb$ structure. The calculated $\Delta E^{pol}$ in GaN is 0.21 and 0.69 eV for 64-atom and 512-atom cells, respectively, which indicates that the 64-atom supercell is too small to provide convergent results. This is due to the fact that in the supercell approach defects induce bands of finite widths, which increase with the decreasing size of the cell. When the widths of the $t_2$-induced bands are larger than the exchange splitting $\Delta\varepsilon_{ex}$ of $t_2$ into the spin-up and spin-down states, the $t_{2\uparrow}$ and $t_{2\downarrow}$ bands overlap. In this case, the calculated spin properties are not correct if the Fermi level in located in the overlap region. In particular, the spin can be lower than 3/2 or 1 for a neutral vacancy in III-V or ZnO crystals.

$V_{Ga}$ in GaN and $V_B$ in BN were previously considered in Ref. [23] within the local spin density approximation. Polarization energies $\Delta E^{pol}$ obtained for neutral $V_{Ga}$ using 216-atom cubic supercells are -0.54 and -0.53 eV for BN and GaN, respectively. This is close to our values but smaller. The difference can be due to the fact that we use larger cells, and vacancy-vacancy coupling (stronger in a smaller cell) is antiferromagnetic [23].

### B. II-VI oxides

We also considered cation vacancies in $w$-ZnO and, to assess the impact of the lattice constant on the properties of vacancies in Sec. V B, in $w$-BeO. The vacancy states are built from the oxygen $p$ orbitals. The triplet $t_2$ is split by the wurtzite crystal field with a small splitting energy $\delta \approx 0.1$ eV. The results are given in Fig. 1d and Table I. In both crystals the $S=1$ high spin state is stable, and indeed this state was observed in ZnO [10]. For ZnO we find $\Delta \varepsilon_{ex}=0.11$ eV and $\Delta E^{pol}=0.04$ eV, while for $V_{Be}$ spin polarization is more pronounced, $\Delta \varepsilon_{ex}=0.44$ eV and $\Delta E^{pol}=0.2$ eV. However, in both crystals the $t_{2\uparrow}$ levels are not resonances but gap states because $\Delta \varepsilon_{ex}$ is smaller than in the nitrides.

High spin state of a neutral $V_{Zn}$ in ZnO was analyzed within GGA using a 128-atom cell [27], and the obtained $\Delta E^{pol}=-0.04$ eV coincides with our result. A four times higher value, $\Delta E^{pol}=-0.16$ eV, was found in [25] using a smaller 64-atom cell. Again, the size of the unit cell and the ferromagnetic character of the $V$-$V$ coupling can explain the difference. As it was pointed out above, small supercells used in Refs. [25,28,29] can lead to non-convergent results and artifacts such as the spin of $V_{Zn}$ lower than 1.

### C. Spin densities

The spatial dependence of the spin density is determined by two factors: the energy of $t_{2\uparrow}$ relative to VBT and the symmetry of the crystal. We begin with the first factor and note that localization of the spin density of $V_{Ga}$ in GaP, or $V_{Si}$ in SiC [1], is strong because $t_{2\uparrow}$ is a deep gap state. A different situation takes place in the III-V nitrides, where $t_{2\uparrow}$ is a resonance degenerate with the valence bands. In this case $t_{2\uparrow}$ hybridizes with the upper part of the valence band, which forces a partial delocalization of its wave function. This difference is highlighted in Fig. 3, where the spin densities of $V_{Ga}$ in GaP (Fig. 3a) and that in $zb$-GaN (Fig. 3b) are compared. In GaP, the spin density is strongly localized on the $sp^3$ orbitals of the vacancy's nearest neighbors. In GaN this contribution is dominant, but the spin density comprises also a long-range tail which involves $p$ orbitals of distant N ions. These orbitals constitute the VBT of GaN. The delocalization of $t_{2\uparrow}$ in turn makes weaker the exchange coupling and reduces the spin polarization. In fact, it is remarkable that in spite of the hybridization the spin polarization of $V_{cation}$ in the nitrides is non-vanishing; this clearly demonstrates robustness of the spin polarization of $p$(N) orbitals.

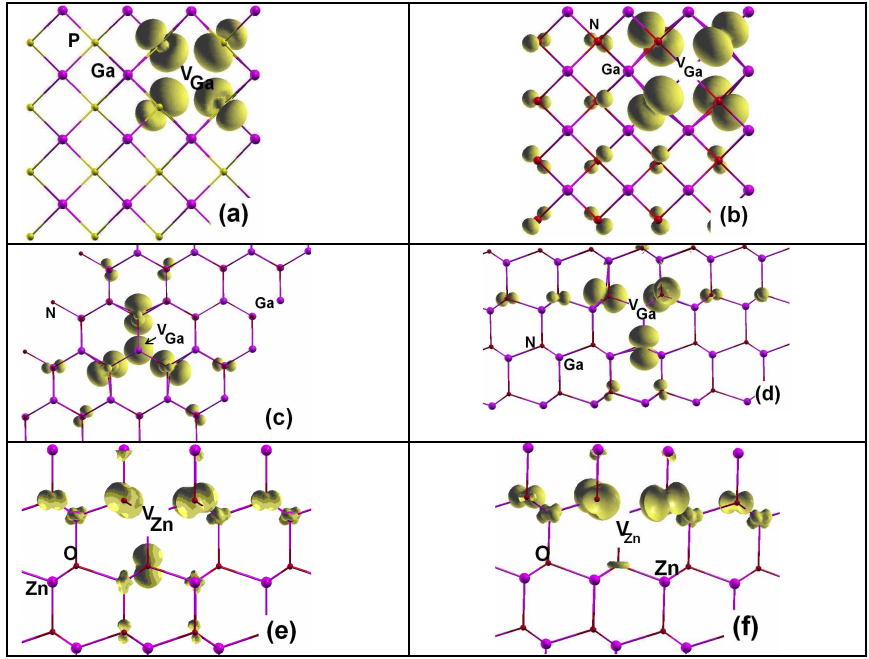

FIG. 3 (color online). Spin densities of (a) $V_{Ga}$ in $zb$-GaP, 64-atom cell, (b) $V_{Ga}$ in $zb$-GaN, 64-atom cell, (c, d) $V_{Ga}$ in $w$-GaN, 128-atom cell; $V_{Zn}$ in $w$-ZnO, 128-atom cell for (e) $q=+1$ and (f) $q=0$ charge states. In the latter case the $t_2$-derived $e_2$ doublet is dominant.

In ZnO, $V_{Zn}$ induces a gap state, and thus spin density of $V_{Zn}$ decays exponentially. However, the decay is rather slow when compared to GaP because of the energetic proximity of $t_{2\uparrow}$ to the VBT, as reflected in Fig.

3e. A characteristic difference with $V_{Ga}$ is the contribution of the $d$(Zn) orbitals to the spin density due to the substantial contribution of $d$(Zn) to the top of the valence band, analogous to the case of shallow acceptors [34]. This contribution is also consistent with Fig. 2g.

Secondly, in the *zb* structure spin polarization has an isotropic character clearly seen in Fig. 3b or in Fig. 3 of Ref. [23]. This contrasts with the anisotropy of the spin polarization in the wurtzite structure, which is apparent from Figs. 2c-f. In this case, the spin density contains the dominant contribution localized on the four vacancy neighbors (like in *zb*-GaN), and four tails extending from the vacancy along the directions of the broken bonds. The first tail is oriented along the *c*-axis, Fig. 3d. The three remaining tails are mainly confined to one atomic (0001) plane (Fig. 3d), and they have a three-fold rotational symmetry and a pronounced directional character displayed in Fig. 3c. The anisotropic character of the spin density of $V_{Zn}$ in ZnO, Figs. 3e and 3f, is very similar to that of $V_{Ga}$ in *w*-GaN, which confirms that the crystal structure is as important as the chemical composition in determining the vacancy properties. Figure 3e holds for the positively charged $V_{Zn}$, with the $t_2\uparrow$ occupied with three electrons. This electronic configuration is identical with that of the neutral $V_{Ga}$, and so are the spin densities (cf. Fig. 3d and 3e). A neutral $V_{Zn}$ has one electron more, which occupies one of the $t_2\downarrow$ levels, namely the $a_1\downarrow$ singlet. Consequently, in the $q=0$ configuration the $a_1$ state is doubly occupied and does not contribute to the spin polarization. Spin density consists in the three quasi-planar tails formed by the $t_2$-derived $e_2\uparrow$ doublet, Fig. 3f.

Finally, we notice a remarkable similarity between the calculated spin density of vacancy states in GaN and ZnO with the vacancy states in graphite observed by scanning tunneling microscopy (STM) [35]. In all these cases the states consist in three highly directional tails extending from vacancy in a quasi-planar geometry, which can indicate that such a shape has a universal character in hexagonal lattices.

## IV. SPIN STATES OF CHARGED VACANCIES

Stability of high spin states depends on the vacancy charge state. This feature is particularly clear in the case of $V_{Ga}$ in GaP, where the high spin state is stable only for the neutral vacancy. In fact, according to the obtained results, adding or subtracting one electron from $t_2$ destabilizes spin polarization, which vanishes because of the relatively weak exchange coupling between the *p* orbitals of phosphorus neighbors. This result can explain why in the experiment only the $S=3/2$ spin state was observed [12]. In the nitrides and oxides the exchange coupling is stronger than in GaP, and it stabilizes the high spin states not only for neutral but also for charged vacancies.

The dependence of the energy levels of $V_{Ga}$ in GaN on the charge state $q$ is given in Fig. 4. With the charge state changing from 0 to 3- the energy of the $t_2$ state increases by ~1 eV due to the increasing intracenter electron-electron repulsion. Next, for $q=1$- and 2- the $t_2\uparrow$ triplet is not a resonance but a deep gap level. The energy of $a_1$ also rises with the increasing occupation of $t_2$: $a_1\downarrow$ is a resonance for $q=0$, while for $q=2$- both $a_1\uparrow$ and $a_1\downarrow$ are in the gap with the spin splitting 0.4 eV, close to that of the triplet. Moreover, in spite of the fact that in the $q=2$- case the total spin is 1/2, the electronic configuration corresponds to the high spin state, since both $a_1$ and $t_2$ are exchange-split. This configuration is analogous to that of a Ni atom with 9 electrons occupying the exchange-split 3$d$ shell. Finally, in the $q=3$- state $V_{Ga}$ assumes the closed shell configuration with vanishing spin.

A second important consequence of the strong exchange coupling in III-N nitrides is that vacancies cannot assume some of the charge states. Namely, $V_{cation}$ in GaN and BN can assume all possible negative charge states, i.e., 1-, 2- and 3-, but the large splitting $\Delta\varepsilon_{ex}$ of $t_2$ leads to an unusual situation where the vacancies cannot achieve positive charge states. This is because $t_2\uparrow$ is a resonance occupied with 3 electrons even in *p*-type samples. The results for one hole per vacancy are referred to as q="1+" charge state for both GaN and BN in Fig. 4, and they show that free holes in the valence band do not appreciably impact the magnetic properties of vacancies. The dependence of the $V_{Al}$-induced levels on the charge state is similar to that of $V_{Ga}$. The energies of both $a_1$ and $t_2$ are somewhat higher than those in GaN. However, in *p*-AlN the singlet $a_1\downarrow$ is in the gap and the vacancy can assume the $q=1+$ state in which electrons occupy both $a_1\uparrow$ and $t_2\uparrow$, the spin-down states are empty, and thus the total spin $S=2$, see Fig. 3b. This is the highest possible spin state of a vacancy, analogous to the $S=5/2$ state of transition metal ions in a crystal. In both cases, the maximum-spin configuration is possible because the strong exchange splitting $\Delta\varepsilon_{ex}$ exceeds the $a_1$-$t_2$ (or $e_2$-$t_2$ in the case of a transition metal impurity) crystal field splitting. The energy levels of $V_B$ in BN are shown in Fig. 3c. The main difference with $V_{Ga}$ and $V_{Al}$ is the fact that in BN even for $q=-2$ charge states a$_1$ is degenerate with the valence band, and therefore the $a_1$-$t_2$ splitting of $V_B$ levels is larger than 1.4 eV.

The case of $V_{Zn}$ is somewhat ambiguous because of the energetic proximity of $t_2$ to VBT. The limited accuracy of the method may affect the final numbers. While for $q=0$ the level is well resolved, it almost merges with the valence band for positively charged vacancies. However, the results given in Table II clearly show that spin polarization persists in *p*-ZnO.

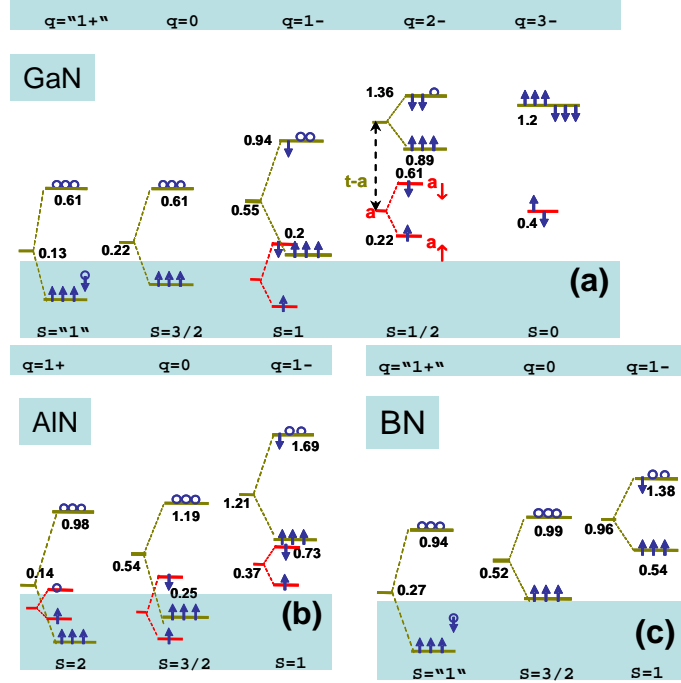

FIG. 4 (color online). Energy level diagram of (a) $V_{Ga}$ in zb-GaN (b) $V_{Al}$ in zb-AlN, and (c) $V_B$ in zb-BN for various charge states, and for "$q=+1$". The numbers give the energies relative to VBT in eV. Energies of the resonance states are shown only schematically. Spins of electrons are indicated by arrows, empty spheres indicate unoccupied states.

TABLE II. Spin polarization energies (in eV) and magnetic moments (in $\mu_B$) for $V_{Zn}$ in ZnO.

| q | 0 | 1+ | 2+ | 3+ |
|---|---|---|---|---|
| $\Delta E^{pol}$ | 0.037 | 0.048 | 0.034 | 0.029 |
| $\mu_{tot}$ | 2.0 | 3.0 | 2.0 | 1.0 |
| $\mu_{abs}$ | 2.53 | 3.54 | 2.62 | 1.97 |

## V. DISCUSSION OF THE RESULTS

### A. Spin polarization energy and the exchange splitting

Figure 5 summarizes the calculated values of $\Delta E^{pol}$ and $\Delta \varepsilon_{ex}$, and shows the correlation between these quantities. The correlation takes place because the energy gain $\Delta E^{pol}$ induced by the formation of the high spin state is partially provided by the energy gain of the electronic contribution to the total energy. This mechanism is illustrated in the inset to Fig. 5. In particular, this simple argument explains why the calculated $\Delta E^{pol}$ is the highest when the $t_2$ state is occupied with 3 electrons, while for other occupations the high-spin state is less stable, or it vanishes like in GaP. Indeed, in the case of $V_{cation}$ in the nitrides, adding one electron reduces $\Delta E^{pol}$ by about one half, Fig. 5, in agreement with this picture. However, this one-electron energy gain accounts for about one half of $\Delta E^{pol}$ since other factors, and in particular the resonant character of $t_{2\uparrow}$ in the nitrides, play an important role.

This argument also accounts for the lower energy of spin polarization of neutral vacancies in ZnO and BeO (where $t_2$ is occupied with 4 electrons) than that in III-V nitrides (where $t_2$ is occupied with 3 electrons). We note, however, that the effect follows mainly from the weaker spin polarization energy of O, 1.53 eV, compared to that of N, 3.19 eV. Similarly, spin polarization energies of As and P atoms (−1.24 and −1.37 eV, respectively) explain the weak stability of $V_{Ga}$ in GaP, and the lack of spin stability in GaAs. This trend exists also in bulk II-V crystals [36].

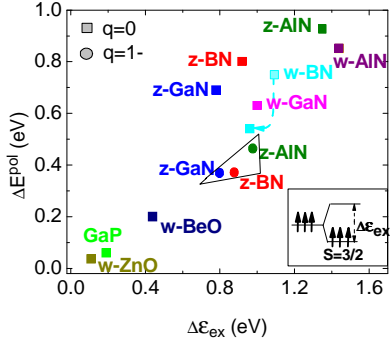

FIG. 5 (color online). Energy of spin polarization of vacancies and exchange splitting of $t_2$ levels. z denotes the *zb* structure. Effect of atomic relaxations is shown for *w*-BN by an arrow. Inset shows the spin splitting of $t_2$ that contributes to polarization energy $\Delta E^{pol}$.

### B. Impact of interatomic distances on the vacancy electronic structure

The interatomic distances between the neighbors of a vacancy are mainly determined by the lattice constant of the host crystal, and also by the atomic relaxations, *i.e.*, displacements of the neighbors. Here, we discuss the influence of both factors on the vacancy's electronic structure. In the studied compounds, the outward relaxation of the nearest neighbors by about 10 %, i.e., 0.2 Å, of the bond length reduces the total energy by the relaxation energy $\Delta E_{relax}$ of 0.6-1.0 eV, see Table III. The displacements of the second and third neighbors are an order of magnitude smaller.

TABLE III. The relaxation energy $\Delta E_{relax}$ and the spin polarization energy $\Delta E^{pol}$ (in eV) in III-N nitrides.

|  | $\Delta E_{relax}$ | | $\Delta E^{pol}$ | |
| --- | --- | --- | --- | --- |
|  | PM | FM | without relaxation | with relaxation |
| $V_{Ga}$ | 0.61 | 0.45 | 0.81 | 0.69 |
| $V_{Al}$ | 0.78 | 0.77 | 0.94 | 0.93 |
| $V_{B}$ | 1.0 | 0.82 | 0.99 | 0.8 |

In the series GaN-AlN-BN, the states of $V_{Ga}$, $V_{Al}$ and $V_B$ are formed by the orbitals of nitrogen neighbors of the vacancy. Thus, the $a_1$-$t_2$ "crystal field splitting" of vacancy states are determined by the overlap of the N dangling bonds given by the N-N interatomic distances. In agreement with this picture, the $a_1$-$t_2$ splitting increases about twice, from 0.75 eV in GaN to 0.95 eV in AlN, and to more than 1.4 eV in BN, and the effect is caused by the decrease in the lattice constant. In the oxides, $a_1$ is deep and not seen.

Spin properties exhibit the same trend: $\Delta E^{pol}$ is 0.81, 0.94, and 0.99 eV for non-relaxed $V_{Ga}$, $V_{Al}$ and $V_B$, see Table II. The outward relaxation of the vacancy neighbors reduces both the overlap between N orbitals and the polarization energy $\Delta E^{pol}$ by ~20 %, as it is shown in Fig. 5 for BN. Considering the oxides, we find $\Delta E^{pol}$ = 0.04 and 0.2 eV for ZnO and BeO, respectively, beceause of the the smaller lattice constant of the latter. These arguments apply also to GaP, where $\Delta E^{pol}$ is an order of magnitude lower than in GaN because both of the weaker atomic spin polarization of P, and of the larger lattice constant of GaP.

From that point of view $V_{Al}$ is an exception, since the impact of relaxation on $\Delta E^{pol}$ is very small. This result is explained by observing that the $a_1\downarrow$ state of $V_{Al}$ is a resonance before the relaxation (similarly to $V_{Ga}$ in GaN), and it becomes a gap state at 0.25 eV above the VBT after the relaxation. The change of character of $a_1$ induces its stronger localization and consequently an increased exchange interaction. These two factors, *i.e.*, the decrease of $\Delta E^{pol}$ caused by the outward displacements of N ions, and its increase caused by the localization of $a_1$, almost cancel each other.

### C. Implications for experiment

*Optical experiments*. The obtained results show in particular that modifications of the electronic structure of $V_{Ga}$ in GaN induced by spin polarization should be observable in experiment. Apart from the fact that the high spin states of vacancies can be identified by electron paramagnetic resonance, we point out that they can be measured in photoluminescence. The characteristic yellow luminescence in GaN is associated with Ga vacancies [37]. One expects that photoluminescence measurements in magnetic fields should reveal polarization dependencies. Moreover, the yellow luminescence line is relatively broad, which may result from the fact that vacancies participating in the luminescence are in various charge states. In other words, the broadening may be due in part to the dependence of the vacancy level on its charge state studied in Sec. IV.

*STM experiments*. The detailed knowledge of the wave function of a defect or dopant in a semiconductor allows identifying it experimentally by STM. This was achieved for, e.g., Zn impurities [38] and magnetic Mn ions [39] in GaAs. Moreover, spin polarization induced by broken bonds of defects was observed by magnetic force microscopy in graphite [3]. The results of Fig. 3 indicate that images induced by a subsurface vacancy in GaN or ZnO depend on the surface polarity. For the (000-1) anion polarity, the STM images are expected to be dominated by the three tails seen in Fig. 3 with a quasi-planar three-fold rotational symmetry, like those seen in graphite [35]. For the (0001) cation polarity the STM image is expected to be more localized, being generated by the fourth *c*-axis oriented tail perpendicular to the surface. A more detailed analysis requires simulations of STM images.

*Collective magnetism*. Following Elfimov *et al.* [7], several theoretical papers on the high spin states of vacancies pointed out that magnetic coupling between vacancies can lead to formation of a collective ferromagnetic phase. Experimentally, ferromagnetism in semiconductors that do not contain magnetic ions was recently observed in GaN nanoparticles [40], ZnO [6], MgO [41, 42], BN [43] and other crystals [41]. Based on theoretical results these authors proposed that the observed FM stems from the presence of cation vacancies.

Here, we make two comments regarding this interpretation. First, in order to give rise to ferromagnetism, the vacancies (or other defects) must be in appropriate charge states, since the vacancy levels must be partially occupied to provide non-vanishing spins. Otherwise, when the defect levels are fully occupied, or when they are empty, their magnetic moments vanish, which obviously excludes formation of any magnetic phase. Accordingly, in $n$-type ZnO and other crystals considered here magnetism induced by $V_{cation}$ vanishes. This point is important since, as a rule, as-grown ZnO is $n$-type with high concentrations of conduction electrons [44]. Magnetism is expected when the Fermi level is situated in the lower part of the band gap. However, in this case the sign of the coupling depends on the vacancy charge state, and is antiferromagnetic for neutral vacancies in the nitrides, and for the $q=+1$ state in ZnO [23]. Finally, in $p$-type nitrides the resonance is occupied with electrons in the high spin state, and a hole-mediated ferromagnetic $V$-$V$ coupling is expected. Concluding, a reliable interpretation of ferromagnetism as a result of the presence of vacancies requires determination of the Fermi level, and is not expected to occur in $n$-type ZnO.

Next, collective magnetism, especially at room temperature, can occur only for sufficiently strong interactions. This condition implies that concentrations of vacancies should exceed equilibrium concentrations by a few orders of magnitude [9]. This is difficult to achieve. For example, the determined concentration of negatively charged $V_{Zn}$ in ZnO [45] is about $10^{14}$ cm$^{-3}$, which is expected to be typical and indeed is orders of magnitude below the percolation threshold for collective magnetism. Given both points, the interpretation of experimental data [40-46] in terms of vacancy-induced effects is open to further studies.

## VI. SUMMARY

In summary, high spin states of cation vacancies were studied for GaAs-GaP-GaN, GaN-AlN-BN and ZnO-BeO series by calculations based on the generalized gradient approximation. This choice is justified by the agreement between the obtained results and the available experimental data. In particular, (i) theory predicts that $V_{Ga}$ in GaP is in the high spin state only in the neutral charge state, in agreement with observations [12], (ii) the $S$=1 spin state of $V_{Zn}$ in ZnO is predicted to be stable, and it was observed [10], and finally (iii) this approach was successfully used to explain the experimentally studied high spin states of vacancies and divacancies in SiC [18].

A systematic analysis reveals several features that help understanding the properties of high spin states, mechanisms stabilizing the spin polarization, and which are referred to experiment. We find in particular that stability of spin polarization of a vacancy increases with the decreasing atomic number of anion in GaAs-GaP-GaN series. This is because the spin polarization energy is determined by the energy of spin polarization of anions, higher for lighter atoms. Similarly, $\Delta E^{pol}$ is higher in the nitrides than in the oxides, since the spin polarization of N is twice higher than that of O.

Strong spin polarization in the nitrides splits the $t_2$ triplet, and the $t_2\uparrow$ partner is a resonance degenerate with the upper valence bands. As a result, vacancies in GaN and BN cannot be positively charged. In $p$-AlN, positively charged $V_{Al}$ is predicted to assume the $S$=2 state, which is the highest possible spin state of a vacancy. In contrast, in ZnO the positively charged vacancies can exist, and they are spin polarized.

Energies of vacancy levels depend on the charge state, and they increase by 0.5-1 eV with the increasing number of electrons occupying $t_2$. This dependence may contribute to the observed broad width of the "yellow luminescence" line in GaN. Furthermore, stability of spin polarization depends on the vacancy charge state. The most stable is the S=3/2 state, when $t_2\uparrow$ is occupied with 3 electrons, and $t_2\downarrow$ is empty. In GaP, this is the only stable (and the only observed) high spin state. In the nitrides, $V_{cation}$ in the 2- charge state has the spin 1/2, but it is in the high spin state with a pronounced $t_2\uparrow$-$t_2\downarrow$ exchange splitting.

Interatomic distances (given by the lattice constant and the atomic relaxations around the vacancy) determine both the singlet-triplet $a_1$-$t_2$ "crystal field splitting", and the stability of the spin polarization.

Finally, wave functions and spin densities of vacancies in a host with the wurtzite structure contain the main part localized on the broken bonds, and four strongly directional long range tails. One is oriented along the $c$-axis, and the three remaining are quasi-planar and perpendicular to the $c$-axis. This feature should be observed by STM.

### ACKNOWLEDGEMENTS

The work was supported by the European Union within European Regional Development Fund, through grants Innovative Economy (POIG.01.03.01-00-159/08, "InTechFun", and POIG.01.01.02-00-008/08, "NanoBiom").

### REFERENCES

∗ Corresponding author, bogus@ifpan.edu.pl.
[1] For a review and references regarding various systems see O. Volnianska and P. Boguslawski, J. Phys. Cond. Matter. **22**, 073202 (2010).

[2] H. Pan, J. B. Yi, L. Shen, R. Q. Wu, J. H. Yang, J. Y. Lin, Y. P. Feng, J. Ding, L. H. Van, and J. H. Yin, Phys. Rev. Lett. **99**, 127201 (2007); H. Wu, A. Stroppa, S. Sakong, S. Picozzi, M. Scheffler, and P. Kratzer, Phys. Rev. Lett. **105**, 267203 (2010).
[3] J. Cervenka, M. I. Katsnelson, and C. F. J. Flipse, Nat. Phys. **5,** 840 (2009).
[4] P. Esquinazi, D. Spemann, R. Höhne, A. Setzer, K.-H. Han, and T. Butz, Phys. Rev. Lett. **91**, 227201 (2003).
[5] Y.-W. Son, M. L. Cohen, and S. G. Louie, *Nature* **444,** 347 (2006).
[6] M. Khalid, M. Ziese, A. Setzer, P. Esquinazi, M. Lorenz, H. Hochmuth, M. Grundmann, D. Spemann, T. Butz, G. Brauer, W. Anwand, G. Fischer, W. A. Adeagbo, W. Hergert, and A. Ernst, Phys. Rev. B **80**, 035331 (2009).
[7] I. S. Elfimov, S. Yunoki, and G. A. Sawatzky, Phys. Rev. Lett. **89**, 216403 (2002).
[8] C. Martínez-Boubeta, J. I. Beltrán, Ll. Balcells, Z. Konstantinović, S. Valencia, D. Schmitz, J. Arbiol, S. Estrade, J. Cornil, and B. Martínez, Phys. Rev. B **82**, 024405 (2010).
[9] J. Osorio-Guillén, S. Lany, S. V. Barabash, and A. Zunger, Phys. Rev. Lett. **96**, 107203 (2006).
[10] D. Galland and A. Herve, Phys. Lett. **33,** A 1 (1970); S. M. Evans, N. C. Giles, L. E. Halliburton, and L. A. Kappers, J. Appl. Phys. **103**, 043710 (2008).
[11] L. E. Halliburton,and D. L. Cowan, W. B. J. Blake and J. E. Wertz, Phys. Rev. B **8**, 1610 (1973); B. H. Rose and L. E. Halliburton, J. Phys. C **7**, 3981 (1974).
[12] T. A. Kennedy, N. D. Wilsey, J. J. Krebs, and G. H. Stauss, Phys. Rev. Lett. **50**, 1281 (1983).
[13] H. Itoh, A. Kawasuso, T. Ohshima, M. Yoshikawa, I. Nashiyama, S. Tanigawa, S. Misawa, H. Okumura, S. Yoshida, Phys. Stat. Sol. (a) **162,** 173 (1997).
[14] N. Mizuochi, S. Yamasaki, H. Takizawa, N. Morishita, T. Ohshima, H. Itoh, T. Umeda, and J. Isoya, Phys. Rev. B **72**, 235208 (2005).
[15] L. Torpo, R. M. Nieminen, K. E. Laasonen, and S. Pöykkö, Appl. Phys. Lett. **74** (2), 221 (1999).
[16] T. Wimbauer, B. K. Meyer, A. Hofstaetter, A. Scharmann, and H. Overhof , Phys. Rev. B **56**, 7384 (1997).
[17] A. Zywietz, J. Furthmüller, and F. Bechstedt, Phys. Rev. B **59**, 15166 (1999).
[18] M. Bockstedte, A. Gali, A. Mattausch, O. Pankratov, and J. W. Steeds, Phys. Stat. Solidi (b) **245**, 1281 (2008).
[19] C. D. Pemmaraju and S. Sanvito, Phys. Rev. Lett. **94**, 217205 (2005).
[20] J. Osorio-Guillén, S. Lany, S. V. Barabash, and A. Zunger, Phys. Rev. B **75**, 184421 (2007).
[21] P. Mahadevan and S. Mahalakshmi, Phys. Rev. B **73**, 153201 (2006).
[22] O. Volnianska, PhD Thesis, Institute of Physics PAS (Warsaw 2009).
[23] P. Dev, Y. Xue, and P. Zhang, Phys. Rev. Lett. **100**, 117204 (2008).
[24] A. Droghetti, C. D. Pemmaraju, and S. Sanvito, Phys. Rev. B **78**, 140404 (R) (2008).
[25] P. Dev and P. Zhang, Phys. Rev. B **81**, 085207 (2010).
[26] F. Wang, Z. Pang, L. Lin, S. Fang, Y. Dai, and S. Han, Phys. Rev. B **80**, 144424 (2009).
[27] H. Peng, H. J. Xiang, S.-H Wei, S.-S Li, J-B Xia, and J. Li, Phys. Rev. Lett. **102**, 017201 (2009).
[28] X. Zuo, S-D. Yoon, A. Yang, W-H. Duan, C. Vittoria, and V. G Harris, J. Appl. Phys, **105,** 07C508 (2009).
[29] T. Chanier, I. Opahle, M. Sargolzaei, R. Hayn, and M. Lannoo, Phys. Rev. Lett. **100**, 026405 (2008).
[30] J. P. Perdew, K. Burke, and M. Ernzerhof, Phys. Rev. Lett.**77**, 3865 (1996).
[31] D. Vanderbilt, Phys. Rev. B **41**, 7892(R) (1990).
[32] www.pwscf.org.
[33] H. J. Monkhorst and J. D. Pack, Phys. Rev. B **13**, 5188 (1976).
[34] O. Volnianska, P. Boguslawski, J. Kaczkowski, P. Jakubas, A. Jezierski, and E. Kaminska, Phys. Rev. B **80,** 245212 (2009).
[35] M. M. Ugeda, I. Brihuega, F. Guinea, and J. M. Gomez-Rodrıguez, Phys. Rev. Lett. **104**, 096804 (2010).
[36] O. Volnianska and P. Bogusławski, Phys. Rev. B **75**, 224418 (2007).
[37] K. Saarinen, T. Laine, S. Kuisma, J. Nissilä, P. Hautojärvi, L. Dobrzynski, J. M. Baranowski, K. Pakula, R. Stepniewski, M. Wojdak, A. Wysmolek, T. Suski, M. Leszczynski, I. Grzegory, and S. Porowski, Phys. Rev. Lett. **79**, 3030 (1997).
[38] R. de Kort, M. C. M. M. van der Wielen, A. J. A. van Roij, W. Kets, and H. van Kempen, Phys. Rev. B **63**, 125336 (2001).
[39] A. M. Yakunin, A.Yu. Silov, P.M. Koenraad, J. H.Wolter, W.Van Roy, J. De Boeck, J.-M. Tang, and M. E. Flatte, Phys. Rev. Lett . **92**, 216806 (2004).
[40] C. Madhu, A. Sundaresan, and C. N. R. Rao, Phys. Rev. B **77**, 201306(R) (2008).
[41] M. Khalid, A. Setzer, M. Ziese, P. Esquinazi, D. Spemann, A. Pöppl, and E. Goering, Phys. Rev. B **81**, 214414 (2010).
[42] M. Kapilashrami, J. Xu, K. V. Rao, L. Belova, E. Carlegrim, and M. Fahlman, J. Phys.: Condens. Matter **22**, 345004 (2010).
[43] B. Song, J. C. Han, J. K. Jian, H. Li, Y. C. Wang, H. Q. Bao, W. Y. Wang, H. B. Zuo, X. H. Zhang, S. H. Meng, and X. L. Chen, Phys. Rev. B **80**, 153203 (2009).
[44] M. D. McCluskey and S. J. Jokela, J. Appl. Phys. **106**, 071101 (2009).
[45] X. J. Wang, L. S. Vlasenko, S. J. Pearton, W. M. Chen, and I. A. Buyanova, J. Phys. D: Appl. Phys. **42**, 175411 (2009).
[46] N. H. Hong, J. Sakai, and V. Brize, J. Phys.: Condens. Matter **19**, 036219 (2007).